\documentclass{elsart}

\usepackage{amsmath}
\usepackage{amsmath}
\usepackage{graphicx}

\begin{document}
\begin{frontmatter}

\title{On the enhancement of nuclear reaction rates in
high-temperature plasma}

\author[KU]{M.~Nakamura\corauthref{cor}},
\corauth[cor]{Corresponding author.}
\ead{nakamura@nucl.kyushu-u.ac.jp}
\author[MU]{V.T.~Voronchev},
\author[KU]{Y.~Nakao}

\address[KU]{Department of Applied Quantum Physics and
Nuclear Engineering, Kyushu University, Hakozaki, Fukuoka
812-0064, Japan}

\address[MU]{Institute of Nuclear Physics, Moscow
State University, Moscow 119992, Russia}

\begin{abstract}
We argue that the Maxwellian approximation can essentially
underestimate the rates of some nuclear reactions in hot plasma
under conditions very close to thermal equilibrium. This
phenomenon is demonstrated explicitly on the example of reactions
in self-sustained DT fusion plasma with admixture of light
elements X = Li, Be, C. A kinetic analysis shows that the
reactivity enhancement results from non-Maxwellian knock-on
perturbations of ion distributions caused by close collisions with
energetic fusion products. It is found that although the fraction
of the knock-on ions is small, these particles appreciably affect
the D+X and T+X reaction rates. The phenomenon discussed is likely
to have general nature and can play role in other laboratory and
probably astrophysical plasma processes.
\end{abstract}

\begin{keyword}
hot plasma \sep nuclear reaction rate \sep ion distribution
function

\PACS 52.55.Pi \sep 25.60.Pj
\end{keyword}
\end{frontmatter}

\section{Introduction}
\label{int}
The concept of nuclear reaction rate is broadly used
in high-temperature plasma research. This rate determines reaction
yield and ultimately specific nuclear power released in plasma.
The yield of a reaction between plasma species 1 and 2 is given by
\begin{equation}\label{yield}
    Y(1{+}2) =
    \alpha n_1n_2 \langle \sigma v\rangle_{12},
\end{equation}
where $\alpha = 1/2$ or 1 for identical or different colliding
nuclei, respectively, $n_1$ and $n_2$ are species densities. The
key quantity in (\ref{yield}) is the reaction rate parameter
$\langle \sigma v\rangle_{12}$ defined as the six-dimensional
integral in velocity space
\begin{equation}\label{reactivity}
    \langle \sigma v\rangle_{12} =
    \int f_1^u(\mathbf v_1)f_2^u(\mathbf v_2)
    \sigma (|\mathbf v_1-\mathbf v_2|)|\mathbf v_1-\mathbf v_2|
    \,d\mathbf v_1 \,d\mathbf v_2.
\end{equation}
Here $\sigma$ is the 1+2 reaction cross section, $\mathbf v_1$ and
$\mathbf v_2$ are particle velocities in the laboratory frame,
$f_1^u$ and $f_2^u$ are unit-normalized particle velocity
distribution functions. In a number of cases the integral
(\ref{reactivity}) can be simplified. For example, in Maxwellian
plasma $\langle \sigma v\rangle_{12}$ takes the well-known form
\begin{equation}\label{Mxw}
    \langle \sigma v\rangle_{12} =
    \left (\frac{8}{\pi \mu} \right )^{1/2}\frac{1}{T^{3/2}}
    \int_0^{\infty} E\sigma (E) \exp \left (-\frac{E}{T}\right )
    \,dE,
\end{equation}
where $\mu$ is the reduced mass of the colliding particles, $E$ is
their kinetic energy in the center-of-mass frame, $T$ is the
plasma ion temperature. The Maxwellian approximation is a
conventional tool to study plasma under conditions close to
thermal equilibrium. At the same time, however, ion distribution
functions in high-temperature plasma strictly speaking are not
purely Maxwellian. A reason of non-Maxwellian deviation lies in
exothermic nuclear reactions proceeding in the plasma. These
reactions generate energetic projectiles which during slowing-down
affect the formation of ion distributions. Charged particles slow
down in the plasma mainly via peripheral (small-angle) Coulomb
scattering by thermal ions and electrons. This mechanism does not
change the equilibrium form of ion distribution. However, since
the energy of reaction products can reach several MeV, close
(large-angle) collisions between them and thermal ions can also
take place in the plasma. The probability of close collisions is
determined by amplitudes of Coulomb and nuclear scattering, and
their interference term. Although such processes occur at rare
opportunity, they can transfer in a single event a large amount of
energy and produce fast knock-on ions. These ions increase the
population of high-energy tails of respective distributions, so
that some deviation from Maxwellian functions appears. Apart from
energetic charged particles, reaction-produced neutrons can also
contribute to the knock-on perturbation mechanism if the plasma is
sufficiently dense.

Since the Maxwellian approximation has widely been used in
laboratory and astrophysical plasma studies, a natural question
arises whether the knock-on perturbation of ion distribution could
in some cases appreciably change reaction rates and affect power
balance in hot plasma systems. It has been recognized that for
conventional DT and DD fusion plasmas the answer is negative. The
combination of three factors -- sizable reaction probabilities at
thermal energies where the majority of ion population is
concentrated, small fractions of knock-on deuterons and tritons,
and moderate energy dependences of the fusion cross sections in
the energy range associated with these fast ions -- makes the
above processes poorly sensitive to slight modifications of D and
T distribution tails. This especially concerns the resonant D+T
reaction whose cross section has a broad maximum at deep
sub-barrier energies. However, in systems composed of nuclei with
$Z>1$ the situation has still been intriguing. In such a system
strong Coulomb repulsion suppresses transmission probability
through the potential barrier between interacting nuclei and, in
the absence of pronounced low-energy resonances, the behavior of
reaction cross section becomes steep at least at sub-barrier
(sub-MeV) energy range. This suggests that the respective reaction
can be sensitive to the form of ion distribution tail, so that the
suprathermal reaction channel induced by knock-on ions may become
appreciable. The purpose of this letter is to investigate the
possible enhancement of nuclear reaction rates due to knock-on
perturbations of ion distributions.

\section{Semi-qualitative consideration}
One can reproduce such situation in a simple two-temperature
model. Let us describe ion distributions in a plasma as
superposition of two functions $f+f'$. The first one is
Maxwellian; it represents the behavior of bulk ions with density
$n$ and temperature $T$. The second function is introduced to
model the ensemble of knock-on ions with density $n'<n$. We assume
that $f'$ also is Maxwellian with some temperature $T'>T$. Then
the total reaction yield (\ref{yield}) can be presented as
\begin{equation}\label{Y12}
    Y(1{+}2) =
    Y_{bulk}\times (1+\lambda),
\end{equation}
where $Y_{bulk}=\alpha n_1n_2 R(T_1,T_2)$ is the thermal yield
provided by bulk particles, while $\lambda$ gives the suprathermal
correction caused by bulk-fast and fast-fast ion interactions
\begin{equation}\label{correc1}
    \lambda =
    \frac{n'_1}{n_1}\frac{R(T'_1,T_2)}{R(T_1,T_2)}
    \frac{1}{\alpha}
    + \frac{n'_2}{n_2}\frac{R(T_1,T'_2)}{R(T_1,T_2)}
    \frac{1}{\alpha}
    + \frac{n'_1n'_2}{n_1n_2}
    \frac{R(T'_1,T'_2)}{R(T_1,T_2)}.
\end{equation}
Here $R$ denotes $\langle \sigma v\rangle$ for the 1+2 reaction
between Maxwellian species with different temperatures.
Substituting Maxwellian distributions with temperatures $\theta_1$
and $\theta_2$ for $f_1^u$ and $f_2^u$ in (\ref{reactivity}), we
find that the two-temperature rate parameter
$R(\theta_1,\theta_2)$ can be reduced to the reactivity
$R(T_{eff})$ for Maxwellian plasma (\ref{Mxw}) with some effective
temperature $T_{eff}$:
\begin{equation}\label{2T-rate}
    R(\theta_1,\theta_2)=R(T_{eff}), \quad
    T_{eff}=\frac{m_2\theta_1 + m_1\theta_2}{m_1+m_2}.
\end{equation}
This allows one to easily estimate $\lambda$. Let us consider, as
an example, different D+X systems: symmetric (X = D), nearly
symmetric and resonant (X = T), asymmetric and involving light
nuclei (X = Li, Be). We assume that only the D distribution is
distorted, while the other particles are Maxwellian. It seems
reasonable to set bulk temperatures of D and X nearly equal,
$T_{\mathrm D}\simeq T_{\mathrm X}=T$, and neglect the
contribution of the fast-fast ion interaction term in
(\ref{correc1}). Under these conditions (\ref{correc1}) takes the
form
\begin{equation}\label{correc2}
    \lambda \simeq
    \frac{n'_{\mathrm D}}{n_{\mathrm D}}
    \frac{R(T'_{\mathrm D},T)}{R(T)}\frac{\beta}{\alpha} ,
\end{equation}
where $\beta$ equals 2 (X=D) or 1 (X$\neq$ D). Choosing $T=10$~keV
typical of fusion plasma level and varying the unknown temperature
$T'_{\mathrm D}$ in the 50--200~keV wide range, we find that the
ratio $R(T'_{\mathrm D},T)/R(T)$ changes approximately within
8--50 (D+D), $5{\times}10^2$--$5{\times}10^3$ (D+$^6$Li),
$10^3$--$5{\times}10^4$ (D+$^7$Li),
$6{\times}10^3$--$7{\times}10^5$ (D+$^7$Be),
$8{\times}10^3$--$6{\times}10^5$ (D+$^9$Be). 
For the D+T reaction this ratio changes from 6.29 to 6.35,
{\it i.e.} proves to be nearly constant.
Such invariant-like behavior with respect to $T'_{\mathrm D}$ results
from resonant nature of the D+T reaction, due to which its reactivity 
rapidly increases at low temperature, peaks around 60~keV and then
becomes essentially insensitive to plasma temperature.
Thus, in the D+Li and D+Be systems a very
small fraction of knock-on deuterons $n'_{\mathrm D}/n_{\mathrm
D}<0.1\%$ makes the contribution of thermal and suprathermal
reaction components comparable, while for the D+T and D+D
reactions the effect is rather imperceptible. One should keep in
mind, however, that these results give approximate picture because
the two-temperature model does not reproduce true form of particle
distributions. Indeed, the knock-on ions are not Maxwellian; at
least their distribution should be truncated at some critical
energy determined by kinematics for particle collision.
Nevertheless, the above estimations indicate that the problem
really stands and it is worth studying rigorously.

\section{Plasma kinetic analysis}
In the present work we employ an appropriate plasma kinetic model
to study the phenomenon on the example of various reactions in
self-sustained DT fusion plasma with admixture of light elements.
The concentration of these elements is assumed to be sufficiently
low to neglect their role when analyzing the behavior of main
plasma species -- fuel ions and 3.5-MeV $\alpha$-particles born in
DT fusions. It was shown (see, for example,
\cite{fish94,ball97,kall00,naka06}) that these energetic
$\alpha$-particles are responsible for non-Maxwellian
perturbations of fuel ion distributions due to $\alpha$-D and
$\alpha$-T close collisions. We describe the behavior of plasma
species $a$ (deuterons, tritons, $\alpha$-particles) with
isotropic velocity distributions in terms of a
Boltzmann-Fokker-Planck (BFP) equation. The BFP equation at
steady-state without external heating can be written in the
following form \cite{naka06,nakao95}:
\begin{equation}\label{BFP}
    \frac{\partial f_a}{\partial t} =
    \left (\frac{\partial f_a}{\partial t}\right )_{Coul.}
    + \left (\frac{\partial f_a}{\partial t}\right )_{NES}
    + \left (\frac{\partial f_a}{\partial t}\right )_{cond.}
    - L_a +S_a =0,
\end{equation}
where $f_a$ is the density-normalized distribution function of
species $a$. The plasma is assumed to satisfy the quasi-neutrality
condition: $n_e = n_d + n_t + 2n_{\alpha}$. The first operator in
the right hand of (\ref{BFP}) represents the effect of small-angle
$a$-ion and $a$-electron Coulomb scattering
\begin{equation}\label{Coulomb}
    \left (\frac{\partial f_a}{\partial t}\right )_{Coul.} =
    \frac{1}{v^2}\frac{\partial}{\partial v}\left
    (A_af_a+B_a\frac{\partial
    f_a}{\partial v}\right ),
\end{equation}
where the functions $A_a$ and $B_a$ are given in
\cite{rose57,kill85,chu85}. The second operator is a Boltzmann
collision integral describing the effect of close $a$-$b$
collision ($b=d,t,\alpha$)
\begin{multline}\label{NES}
    \left (\frac{\partial f_a}{\partial t}\right )_{NES}
    =\sum _b \frac{2\pi}{v^2}\int_0^\infty v'f_a(v')
    \int_0^\infty v_b'f_b(v_b')P(v'\rightarrow v|v_b) \\
    \times
    \left (\int_{|v'-v'_b|}^{v'+v'_b} v_r'^2\sigma _{NES}
    (v'_r) dv'_r\right )dv'dv'_b
    -\sum _b \frac{2\pi}{v}f_a(v)
    \int_0^\infty v_bf_b(v_b) \\
    \times
    \left (\int_{|v-v_b|}^{v+v_b} v_r^2\sigma _{NES}
    (v_r) dv_r\right )dv_b.
\end{multline}
Here $v'_r=|v'-v'_b|$, $v_r=|v-v_b|$, $P(v'\rightarrow v|v_b)$
gives the probability distribution function for the speed $v$ of a
scattered particle, and $\sigma _{NES}$ is the collision cross
section quoted from \cite{perk81}. The third term
\begin{equation}\label{cond}
    \left (\frac{\partial
    f_a}{\partial t}\right )_{cond.} =
    \frac{1}{v^2}\frac{\partial}{\partial v}\left
    (\frac{v^3f_a}{2\tau _C^{(a)}(v)} \right )
\end{equation}
gives the diffusion in velocity space due to thermal conduction
with the typical time $\tau _C^{(a)}$. Finally, $L_a$ and $S_a$
are particle loss and source terms, respectively, taking different
forms for every ion species. Plasma electrons are considered to be
Maxwellian at some temperature $T_e$ incorporated in our model by
using a global power-balance (GPB) equation. This equation
determines the relation between plasma density and temperature,
and has the simple form\begin{equation}\label{GPB}
    P_{heat}(n_i,T_i)
    -P_{brem}(n_i,n_e,T_e)-P_{c.p.}(n_i,n_e,T_i,T_e)=0.
\end{equation}
Here $P_{heat}$ is the plasma heating rate by $\alpha$-particles,
$P_{brem}$ is the rate of bremsstrahlung energy loss, $P_{c.p.}$
gives the energy loss due to thermal conduction and particle leak.
At chosen $T_e$, being an input parameter in our model, plasma
density is estimated from (\ref{GPB}) assuming $n_e\simeq n_d+n_t$
and $T_e\simeq T_i$. The detailed description of the kinetic model
and explicit expressions for all terms in the BFP and GPB
equations one can find in \cite{naka06}.

\begin{figure}
\begin{center}
\includegraphics*[width=3.2in]{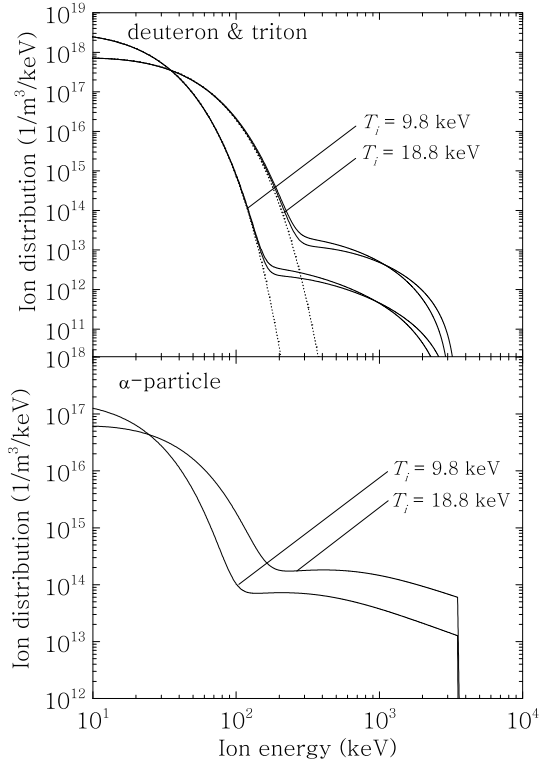}
\end{center}
\caption{\label{fig:distr} The energy distribution
functions of fuel ions and $\alpha$-particles calculated under two
plasma conditions: (i) $T_e = 10$~keV, $T_i = 9.8$~keV, $n_e =
1.2\times 10^{20}$~m$^{-3}$, $n_d = n_t = 5.8\times
10^{19}$~m$^{-3}$, and (ii) $T_e = 20$~keV, $T_i = 18.8$~keV, $n_e
= 6.1\times 10^{19}$~m$^{-3}$, $n_d = n_t = 2.8\times
10^{19}$~m$^{-3}$. The respective Maxwellian distributions are
shown by the dotted curves.}
\end{figure}

Figure~1 shows the particle distributions in energy space
calculated under conditions close to ITER-like plasma. The
non-Maxwellian perturbations of the deuteron and triton functions
caused by $\alpha$-D and $\alpha$-T close collisions are clearly
marked at energies above a few hundred of keV.\footnote{We note
that knock-on deuterons were already observed in DT fusion
experiments at JET \cite{koro00}.} The distribution of
$\alpha$-particles reflects well slowing down history of these
fusion products. This distribution reveals moderate energy
dependence in the 0.1--3.5~MeV deceleration range, while at
thermal energies it is described by Maxwellian-like form. The
plasma ion temperature $T_i$ in Fig.~1 is evaluated as
\begin{equation}\label{Ti}
    T_i =
    \left(n_d^{bulk}T_d+n_t^{bulk}T_t\right )/
    \left(n_d^{bulk}+n_t^{bulk}\right ),
\end{equation}
where the temperatures of deuterons $T_d$ and tritons $T_t$ are
obtained by fitting the bulk components of the D and T
distributions to proper Maxwellian functions. 
For conditions considered in the work the difference between
$T_d$ and $T_t$ does not exceed 4~\%.
Although it is rather small, we employ the general difinition of
$T_i$ Eq.~(\ref{Ti}) instead of its reduction $T_d=T_t=T_i$
to estimate this basic plasma parameter as accurately as possible.
An informative
parameter is the fraction of the knock-on ions $n'/n$. It is
estimated to be at the level of 0.03\,\% indicating that the
plasma conditions are very close to thermal equilibrium.

Now we can examine the influence of the knock-on ions on reaction
rates in the DT/X plasma. In the present study the admixture ions
X are chosen to be Li, Be, C. These light elements are often
considered as low-$Z$ impurity in magnetic confinement fusion
devices. For example, Li was already used in operation of TFTR
\cite{mans01}, ASDEX and TEXTOR tokamaks, W-7 AS stellarator
\cite{bran99,schw92}, and has been proposed as a diagnostic
admixture for fusion machines of next generation
\cite{voro01,voro03,naka04}. Be and C have been vigorously used
for plasma diagnostics in JET \cite{kipt04,mant02}. In order to
make the study most informative we examine the variety of
reactions having different mechanisms. They are listed in
Table~\ref{table1} including some processes proposed for plasma
$\gamma$-ray spectroscopy, the energy-producing $\mathrm
Li(d,\alpha)$ and tritium-breeding $\mathrm Li(d,pt)$ reactions,
the conventional D+T and D+D fusion processes. The respective
reaction cross sections are plotted in Fig.~2.

\begin{table}
\caption{\label{table1}The list of nuclear reactions in the DT/X
plasma and the enhancement of their rate parameters caused by the
knock-on deuterons and tritons}
\begin{tabular*}{\hsize}{lllll}
\hline system~~~~~~ & reaction~~~~~~~~~~~~~~~~ & $Q$-value~~~~~~ &
$E_{\gamma}^{\mathrm a}$~~~~~~~~~~~ & $\langle \sigma v\rangle /
\langle \sigma v\rangle _{\mathrm {Mxw}} $ \\
 & & (MeV) & (MeV) & $T_i=10$--40~keV \\
\hline D+Li & $^6\mathrm{Li}(d,n_1)^7\mathrm{Be}^{\ast}$ & 2.95 &
0.429 &
2.2--1.5 \\
 & $^6\mathrm{Li}(d,p_1)^7\mathrm{Li}^{\ast}$ & 4.55 & 0.478 &
2.5--1.5 \\
 & $^6\mathrm{Li}(d,pt)\alpha$ & 2.56 & & 2.9--1.6 \\
 & $^6\mathrm{Li}(d,\alpha)\alpha$ & 22.37 & & 1.5--1.2 \\
T+Li & $^6\mathrm{Li}(t,d_1)^7\mathrm{Li}^{\ast}$ & 0.51 & 0.478 &
7.8--1.7 \\
 & $^6\mathrm{Li}(t,p_1)^8\mathrm{Li}^{\ast}$ & - 0.18 & 0.981 &
10$^8$--20 \\
D+Be & $^9\mathrm{Be}(d,\gamma)^{11}\mathrm{B} $ & 15.81 & &
50--3 \\
D+C & $^{12}\mathrm{C}(d,p_1)^{13}\mathrm{C}^{\ast}$ & - 0.37 &
3.089 & 10$^{18}$--10$^4$ \\
D+T & $\mathrm{D}(t,n)\alpha$ & 17.59 & & $\leq$ 1.01 \\
D+D & $\mathrm{D}(d,n)^3\mathrm{He}$ & 3.27 & & $\leq$ 1.07 \\
\hline
\end{tabular*}
$^{\mathrm a}$ Energies of $\gamma$ rays emitted by the excited
daughter nuclei
\end{table}

\begin{figure}
\begin{center}
\includegraphics*[width=3.2in]{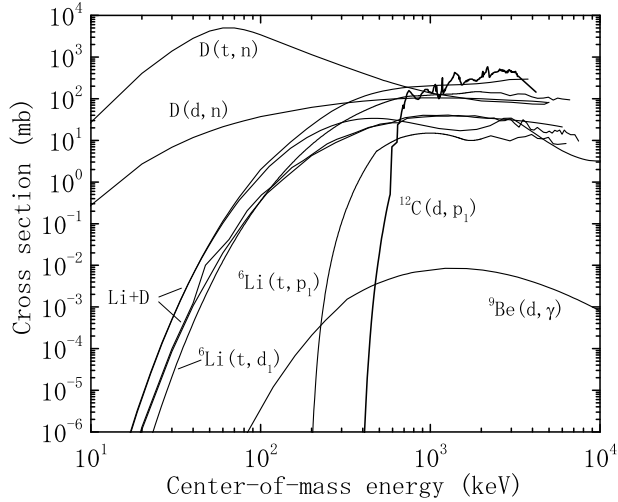}
\end{center}
\caption{\label{fig:sigma}The cross sections of reactions listed
in Table~1. The D+Li curves are not resolved well at sub-barrier
energies.}
\end{figure}
\begin{figure}
\begin{center}
\includegraphics*[width=3.2in]{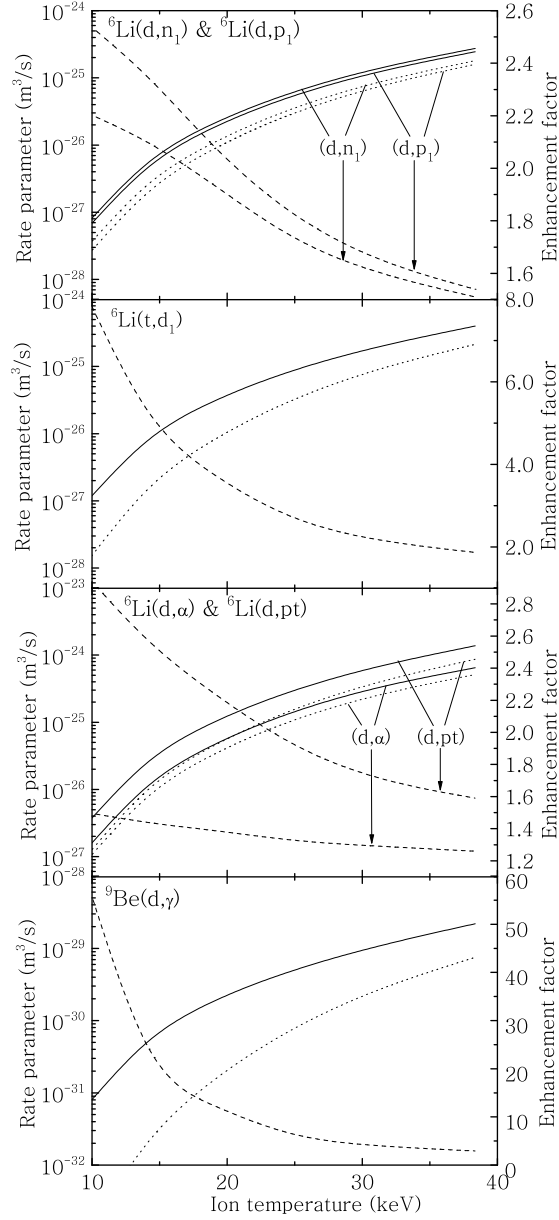}
\end{center}
\caption{\label{fig:rate}The rate parameters for exothermic
reactions from Table~1. The computed and Maxwellian data are shown
by the solid and dotted curves, respectively. The dashed curves
give the factor $\delta =\langle \sigma v\rangle / \langle \sigma
v\rangle_{\mathrm {Mxw}}$.}
\end{figure}
\begin{figure}
\begin{center}
\includegraphics*[width=3.2in]{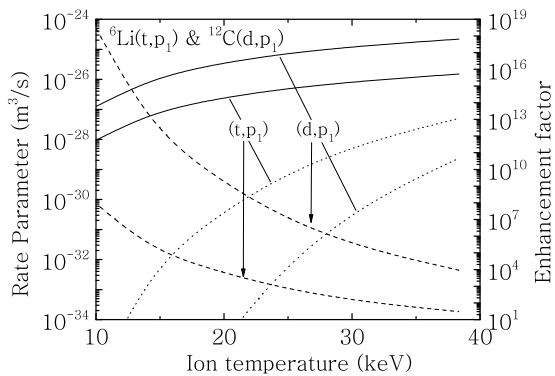}
\end{center}
\caption{\label{fig:rate}The rate parameters for endothermic
reactions from Table~1. }
\end{figure}

To calculate $\langle \sigma v\rangle$ for the distorted isotropic
distributions of D and T, we reduce the general expression for
reactivity (\ref{reactivity}) to the form
\begin{equation}\label{react12}
    \langle \sigma v\rangle_{12} =
    \frac{8\pi^2}{n_1n_2} \int_0^{\infty} v_1f_1(v_1)
    \int_0^{\infty} v_2f_2(v_2)
    \int_{|v_1-v_2|}^{v_1+v_2} v^2\sigma (v)
    \,dv\,dv_1\,dv_2,
\end{equation}
where $v$ is the relative speed $|\mathbf v_1-\mathbf v_2|$.
Assuming the comparatively heavy particles X to be Maxwellian, the
reaction rate parameters have been computed at ion temperature
$T_i=10$--40~keV. The results are plotted in Figs.~3 and 4, and also
displayed in Table~\ref{table1}. We see that although the fraction
of the knock-on ions is only 0.03\,\%, these particles appreciably
affect the D+X and T+X reactivity. It is underestimated if
Maxwellian DT plasma is assumed, and for some reactions the
discrepancy between the two approaches becomes crucial. The ratio
$\delta =\langle \sigma v\rangle / \langle \sigma
v\rangle_{\mathrm {Mxw}}$ monotonically increases with decreasing
$T_i$ down to the plasma ignition point, and for the exothermic
reactions presented in Fig.~3 it changes approximately within
1.5--2.5 (D+Li), 2--8 (T+Li) and 3--50 (D+Be). Special attention
is worth being paid to the endothermic T+Li and D+C reactions
displayed in Fig.~4. Here $\delta$ turns out to be several orders
of magnitude or more, that reflects threshold nature of these
processes. Both of them are forbidden at energies below
thresholds, so only sufficiently fast ions contribute to the
reactions. At the same time, Fig.~1 shows that the amount of these
particles is essentially underestimated in Maxwellian plasma. The
record enhancement is marked for D+C; the high threshold and
strong Coulomb suppression of the thermal channel make the role of
the knock-on deuterons extremely important here. Thus, the both
reactions proceed via suprathermal channels and are solely
governed by knock-on ions. This may have interesting applications
in fusion technology. For example, 0.981-MeV photons emitted in
$^6\mathrm Li(t,p_1)$ might be applicable to energetic triton and
$\alpha$-particle diagnostics \cite{naka06}.

In agreement with the comments in Section~\ref{int} we find that
the knock-on ions do not significantly affect the D+T and D+D
reactions. Table~\ref{table1} shows that the enhancement factor
$\delta$ does not exceed 1\,\% (D+T) and 7\,\% (D+D).

\section{Conclusion}
We have presented arguments that the Maxwellian approximation can
essentially underestimate reaction rates in nearly thermal-equilibrium
plasma, and explicitly demonstrated this phenomenon for some
reactions in the DT/X plasma. The enhancement of reactivity
results from the knock-on perturbations of ion distributions
caused by close collisions with energetic fusion products. The
numerical analysis carried out in the work is consistent with the
semi-qualitative consideration, but the level of reactivity
enhancement marked has turned out to be surprisingly high.

The phenomenon is likely to have general nature and can play role in
other plasma systems. This especially concerns threshold nuclear
processes -- our study indicates that evaluation of their rates
within the Maxwellian approximation can involve dramatic errors
and prove to be fully useless. It seems possible that knock-on
ions can affect power balance and even reduce ignition temperature
for some exotic fuels, in which conditions favorable for the
non-Maxwellian pumping of ion distributions can be realized. The
aneutronic $^3$He plasma would be an interesting object for such a
study. Indeed, the $^3$He+$^3$He reaction has large $Q=12.9$~MeV,
generates fast charged particles and exhibits an appropriate cross
section behavior. The reaction cross section rapidly and
monotonically rises with increasing energy up to the MeV region.

Apart from laboratory plasmas, the phenomenon discussed may also
appear in some astrophysical processes. Primordial plasma is of
particular importance here. Standard big-bang nucleosynthesis
(BBN) relies on nuclear reaction network involving Maxwellian
reactivity, and accuracy of nuclear inputs has been under
attention \cite{cybu04,noll00}. If the mechanism of non-Maxwellian
deviation would come into play under BBN specific conditions, it
could update input reactivities and thereby offer new insight into
synthesis of light elements in the early universe. Analysis of
this scenario requires coupled cosmological and plasma kinetic
calculations which were beyond the scope of our work. However, the
demonstration that a very small, almost negligible fraction of
knock-on particles ($\sim 10^{-4}$) in hot plasma can
significantly change the rates of some reactions gives impetus to
such study.

\section*{Acknowledgments} 
M.N. and Y.N. would like to thank
Dr.~H.~Matsuura for fruitful discussions. The financial support of
Research Fellowship of the Japan Society for the Promotion of
Science for Young Scientists is acknowledged by M.N.

\end{document}